\documentclass[showpacs,amsmath,amssymb,onecolumn,prx,superscriptaddress,nofootinbib]{revtex4-1}
\usepackage[dvips]{graphicx} % for figures
\usepackage{amsfonts}
\usepackage{amssymb}
\usepackage{amscd}
\usepackage{amsmath}    % need for subequations
\usepackage{enumerate}
\usepackage{epsfig}
\usepackage{subfigure}
\usepackage{xcolor}
\usepackage{amsthm}
\usepackage{slashbox}

\newcommand{\bra}[1]{\mbox{$\left\langle #1 \right|$}}
\newcommand{\ket}[1]{\mbox{$\left| #1 \right\rangle$}}
\newcommand{\braket}[2]{\mbox{$\left\langle #1 | #2 \right\rangle$}}

\def\tr{{\rm tr}}

\begin{document}

%\title{Unmatched-basis data can enhance the security of measurement-device-independent quantum key distribution}
\title{Mismatched-basis statistics enable quantum key distribution with uncharacterized qubit sources}

%%Mismatched-basis statistics enable quantum key distribution with uncharacterized qubit sources and measurements

\author{Zhen-Qiang Yin}
\address{Key Laboratory of Quantum Information, University of Science and Technology of China, Hefei 230026, China\\
and Synergetic Innovation Center of Quantum Information $\&$ Quantum Physics, University of Science and Technology of China,\\
Hefei, Anhui 230026, China}
\author{Chi-Hang Fred Fung}
\email{chffung@hku.hk}
\address{Department of Physics and Center of Theoretical and Computational Physics, University of Hong Kong, Pokfulam Road, Hong Kong}
\author{Xiongfeng Ma}
\email{xma@tsinghua.edu.cn}
\address{Center for Quantum Information, Institute for Interdisciplinary Information Sciences, Tsinghua University, Beijing, China}
\author{Chun-Mei Zhang}
\address{Key Laboratory of Quantum Information, University of Science and Technology of China, Hefei 230026, China\\
and Synergetic Innovation Center of Quantum Information $\&$ Quantum Physics, University of Science and Technology of China,\\
Hefei, Anhui 230026, China}
\author{Hong-Wei Li}
\author{Wei Chen}
\email{kooky@mail.ustc.edu.cn}
\author{Shuang Wang}
\email{wshuang@ustc.edu.cn}
\author{Guang-Can Guo}
\affiliation{Key Laboratory of Quantum Information, University of Science and Technology of China, Hefei 230026, China\\
and Synergetic Innovation Center of Quantum Information $\&$ Quantum Physics, University of Science and Technology of China,\\
Hefei, Anhui 230026, China}
\author{Zheng-Fu Han}
\affiliation{Key Laboratory of Quantum Information, University of Science and Technology of China, Hefei 230026, China\\
and Synergetic Innovation Center of Quantum Information $\&$ Quantum Physics, University of Science and Technology of China,\\
Hefei, Anhui 230026, China}

\begin{abstract}
In the postprocessing of quantum key distribution, the raw key bits from the mismatched-basis measurements, where two parties use different bases, are normally discarded. Here, we propose a postprocessing method that exploits measurement statistics from mismatched-basis cases, and
%which takes advantage of the measurement outcome statistics from mismatched-basis cases.
prove that
%This method
incorporating these statistics enables uncharacterized qubit sources to be used in the measurement-device-independent quantum key distribution protocol and the Bennett-Brassard 1984 protocol, a case which is otherwise impossible.

%The method is applied to the measurement-device-independent QKD protocol and show the power of mismatched-basis statistics.

%Measurement-device-independent quantum key distribution (MDIQKD) defies all side-channel attacks at detection.
%%can remove any possible detection side channels.
%We propose an improvement on our previous qubit-MDIQKD scheme, sharing the same capability of allowing uncharacterized encoding systems to be used as qubit sources, but with an enhanced key generation rate.
%This is made possible by using the measurement statistics of qubits from mismatched bases.
%Our new scheme also allows a simple detection structure because only projection onto one Bell state is needed.
%We prove the security and perform simulation using realistic parameters to show that our new scheme achieves practical key rates.
%Also, our analysis is directly applicable to the BB84 protocol where uncharacterized qubit sources and measurements are used.

%is practical.
%In original proposal, the legitimate communicating peers will perform basis reconciliation, in which the key bits corresponding to mismatched basis are discarded and only those corresponding to matched basis are retained. Here, we find that the statistics for mismatched basis may be useful to enhance the security of MDIQKD. The statistics for mismatched basis cases can help Alice and Bob to relax their assumptions on the characteristics on encoding states. Simulation results show that the new scheme is practical.
\end{abstract}


\maketitle

\section{introduction}
Quantum key distribution (QKD) \cite{Bennett:BB84:1984,Ekert:QKD:1991} exploits quantum mechanical effects to generate secret keys between two users, Alice and Bob, against a quantum eavesdropper, Eve.
Such a key may then be used to encrypt further communications between Alice and Bob using the one-time pad which has been proven to be information-theoretically secure by Shannon \cite{shannon1949communication}.
QKD has also been proven to be secure.
%
%In 1949, one-time pad encryption has been proven to be information-theoretically secure by Shannon \cite{shannon1949communication}, which requires the private key to have the same length as the message. Hence, the secure communication task becomes a question of how to distribute secure keys.
%%do secure key distribution (or extension).
%Quantum key distribution (QKD) \cite{Bennett:BB84:1984,Ekert:QKD:1991} solves this problem by providing a means of extending secret keys between two users, Alice and Bob, against a quantum eavesdropper, Eve.
Initial security proofs of QKD
%Initial efforts in proving the security of QKD
focused on the situation that
%The security of QKD is proven to be secure when
trusted or well-characterized devices are used \cite{Mayers:Security:2001,Lo:QKDSecurity:1999,Shor:Preskill:2000}.
%In the meantime,
%Meanwhile,
Furthermore,
many QKD experiments have also been successfully demonstrated \cite{Zhao:DecoyExp:2006,Rosenberg:ExpDecoy:2007,Zeilinger:ExpDecoy:2007,Peng:ExpDecoy:2007, Zhao:Decoy60km:2006,Yuan:ExpDecoy2007,Takesue:40dBQKD:2007,Stucki:250kmQKD:2009, Wang:260kmQKD:2012,Frolich:QKDnet:2013}.
However,
the applicability of these initial proofs in real-life situations is questionable, since
%in reality,
realistic
devices can be untrusted or uncharacterized because they may be manufactured by Eve or they simply operate imperfectly.

The security problem caused by using untrusted devices
is a real problem, as
%has been
demonstrated by
various hacking strategies on practical QKD systems, including the fake-state attack \cite{MAS_Eff_06,Makarov:Fake:08}, time-shift attack \cite{Qi:TimeShift:2007,Zhao:TimeshiftExp:2008}, phase-remapping attack \cite{Fung:Remap:07,Xu:PhaseRemap:2010}, detector-blinding attack \cite{Lydersen:Hacking:2010,Gerhardt:Blind:2011}, and unambiguous state discrimination (USD) attack \cite{Tang:USD:2013}. In these attacks, device imperfections in QKD systems are exploited. From the study of hacking, we learn that the major security issues lie in the detection system. Hence, how to remove detector side channels becomes a key question in the area. To solve this problem, QKD protocols that are secure against detection loopholes have been proposed \cite{MXF:DDIQKD:2012,Pawlowski:Semi:2011}, none of which, however, is practical. Lo, Curty, and Qi presented a seminal work of measurement-device-independent QKD (MDIQKD) \cite{Lo:MDIQKD:2012}, which can be practically implemented and is immune to all possible detector side channel attacks. Recently, several experimental demonstrations of MDIQKD prove its practicality \cite{Rubenok:MIQKDexp:2013,liu:MIQKDexp:2013,daSilva:MIQKD:2013,Tang:MDIQKDexp:2013}.
The MDIQKD scheme shares the advantage with the BB84 protocol \cite{Bennett:BB84:1984} that no entanglement is needed. Another approach that solves all side channel problems at both source and receiver is by using an entanglement source in the device-independent QKD (DIQKD) scheme~\cite{Acin:DeviceIn:07,Pironio:DeviceIn:09,Branciard:OneDIQKD:2012,Lim:DIQKD:2013}.
%To solve the security problem in the case of untrusted or uncharacterized devices, two main QKD schemes have been proposed.

MDIQKD and BB84 are attractive schemes for practical implementations because of their long achievable distances and they operate in the prepare-and-measure manner.
%, and they are the focus of this paper.
DIQKD suffers from the need of low loss channels and detectors, limiting the distance, and the use of entanglement.
But at this cost,
%even though
DIQKD is superior in that it allows the source and measurements to be completely uncharacterized.
In contrast, a major common problem of standard MDIQKD and BB84 is that they require the source states to be perfect, or well characterized~\cite{Tamaki:MIQKD:2012, Wang:MDIQKD:2013}.
Otherwise, if the source states can be arbitrary, it can be easily shown that they cannot generate any secret key.
In this paper, we prove that by incorporating the mismatched-basis data in the security analysis, MDIQKD and BB84 can generate secret keys even when the source states are uncharacterized qubits.
This is a modification to
standard MDIQKD and BB84 which discard mismatched-basis data and ignore their statistics.
In essence, our method endows MDIQKD and BB84 with a higher level of device independency, approaching that of DIQKD.
We note that our method still requires the source states to be qubits while this is not necessary in DIQKD.
On the other hand,
we remark that our qubit assumption is not too stringent in many practical MDIQKD and BB84 systems.
%Note that our assumptions satisfy many practical MDIQKD and BB84 systems.
For example, in phase encoding systems, it is reasonable to assume that the encoding states are in two-dimensional space while the accuracies of the phase modulators may be questionable.

%Our postprocessing contrasts with standard MDIQKD and BB84 which discard mismatched-basis data and ignore their statistics.

We remark that using the mismatched-basis statistics in security analysis has been proposed before.
%~\cite{Barnett:mismatchedbasis:1993,Watanabe:mismatchedbasis:2008}.
Barnett {\it et al.}~\cite{Barnett:mismatchedbasis:1993} showed that mismatched-basis statistics alone can detect the presence of intercept-and-resend attacks by Eve when perfect source states are used.
Watanabe {\it et al.}~\cite{Watanabe:mismatchedbasis:2008} used these statistics to improve the key generation rate of BB84 for some types of channels but perfect source states are still assumed.
Recently, Tamaki {\it et al.}~\cite{Tamaki:mismatchedbasis:2013} provided a scheme that uses these statistics to mitigate the adverse effect of source errors but it requires full characterization of the imperfect source qubit states.
Here, our work is very different; we use the mismatched-basis statistics to lift some restriction on the source. No detailed characterization of the qubit source is needed.
Essentially, we show that the case of no security at all (where the source qubit states are uncharacterized) can be made secure by using the mismatched-basis statistics.
%We also note that a related and independent work on using mismatch-basis statistics to mitigate source errors has recently been studied by Tamaki {\it et al.}~\cite{Tamaki:mismatchedbasis:2013}.

%In closer detail,
Let us look into the issue of mismatched basis in more detail.
In the BB84 protocol, the encoding states of Alice are the eigenstates of the Pauli operators $Z$ or $X$, while Bob performs the $Z$- or $X$-basis measurements
%projections
randomly to measure the quantum state
%of the photons
sent by Alice.
%The selection of $Z$ or $X$ is called basis of Alice or Bob.
In standard BB84, only the key bits and statistics of matched-basis cases (i.e., Alice and Bob choose the same basis) are considered while the cross-basis data are discarded. This is reasonable
%for the original BB84 protocol,
since Alice completely knows her encoding states and Bob is also sure that his measurement is either $Z$ or $X$.
% projection.
Thus, the statistics of mismatched basis are not needed in general.
% for Alice and Bob.
%Thus, the statistics of mismatched basis are not useful for Alice and Bob.
However, when we consider that Alice's encoding operations and Bob's measurements are not fully characterized, the statistics for mismatched basis
are needed.
%may be not useless.
As an example, we consider that Alice's encoding states are all eigenstates of $Z$ and Bob's measurements are all $Z$ projections.
%, due to some unexpected reason.
If Alice and Bob are unaware of that, the protocol is of course not secure. But Alice and Bob can exclude this error if they observe the statistics of the mismatched basis. Hence, the mismatched basis statistics should help the QKD protocol to be secure even when there are some imperfections in their devices.
We provide a proof for this in this paper.

%%In reality, device imperfections may lead to security loopholes in QKD systems. Recently, hacking strategies on practical QKD systems have been proposed, such as fake-state attack \cite{MAS_Eff_06,Makarov:Fake:08}, time-shift attack \cite{Qi:TimeShift:2007,Zhao:TimeshiftExp:2008}, phase-remapping attack \cite{Fung:Remap:07,Xu:PhaseRemap:2010}, detector-blinding attack \cite{Lydersen:Hacking:2010,Gerhardt:Blind:2011}, and unambiguous state discrimination (USD) attack \cite{Tang:USD:2013}. In these attacks, device imperfections in the measurement device of QKD systems are exploited. From the study of hacking, we learn that the major security issues lie on the detection system. Hence, how to remove detector side channels becomes a key question in the area. To solve this problem, QKD protocols that are secure against detection loopholes have been proposed \cite{MXF:DDIQKD:2012,Pawlowski:Semi:2011}, none of which, however, is practical. Lo, Curty, and Qi presented a seminal work of measurement-device-independent QKD (MDIQKD) \cite{Lo:MDIQKD:2012}, which can be practically implemented and is immune to all possible detector side channel attacks (see also Ref.~\cite{Braunstein:MIQKD:2012}). Recently, several experimental demonstrations of MDIQKD prove the its practicality \cite{Rubenok:MIQKDexp:2013,daSilva:MIQKD:2013,liu:MIQKDexp:2013,Tang:MDIQKDexp:2013}.

In the original MDIQKD protocol, Alice and Bob each encode their traveling qubits randomly from $\{\ket{0}, \ket{1}, |+\rangle, |-\rangle\}$, and send them to a measurement unit (MU) controlled by an untrusted party Eve, who is supposed to perform a Bell-state measurement (BSM) on the incoming qubit pairs. Eve announces a message to Alice and Bob according to her measurement result. A secure key can then be established between Alice and Bob given Eve's announcements.
% of the BSM results.
The advantage of MDIQKD is that its security does not rely on any assumption of the MU, which can even be assumed to be fabricated or controlled by Eve; also, Eve is allowed to not cooperate and lie.  Even under these settings, the final key is still secure.  However, the security of MDIQKD relies on the assumption that Alice and Bob are able to characterize
%the accuracy of
their encoding systems \cite{Tamaki:MIQKD:2012, Wang:MDIQKD:2013}. Recently, by modifying the original MDIQKD and assuming qubit sources, we have proved that even when the
%accuracies of
encoding systems are totally unknown, MDIQKD can still be secure \cite{Yin:QMDIQKD:2013}. In this modified MDIQKD scheme which we call qubit-MDIQKD \cite{Yin:QMDIQKD:2013}, the MU must be able to distinguish two Bell states, while
the original MDIQKD protocol identifies only one Bell state.
%in the original MDIQKD protocol, it is sufficient for the MU to identify only one Bell state.

In this paper,
we propose a modification to the original MDIQKD protocol.
Unlike qubit-MDIQKD, our new scheme only needs to identify one Bell state, while still allowing uncharacterized qubit encoding systems, thanks to the incorporation of the mismatched-basis statistics.
%we propose an improvement on the qubit-MDIQKD scheme that identifies only one Bell state, while still allowing uncharacterized encoding systems.
Here, we prove the security of this new scheme and show that it outperforms the qubit-MDIQKD scheme.
Our main proof here is for the new MDIQKD scheme and we can specialize it to work on BB84 as well.
The idea is to regard the MU and Bob in qubit-MDIQKD as Bob in BB84.
Thus, with one proof, we cover the security of both MDIQKD and BB84 using uncharacterized qubit sources.
In BB84, our proof also allows uncharacterized qubit von Neumann measurements to be used.

The rest of the article is organized as follows. In Sec.~\ref{main result}, we present the details of the proposed MDIQKD scheme and the main result of our security proof.
%, and main result of our security proof is also introduced here.
The details of our security proof are given in the appendix.
We adapt our analysis to BB84 in Sec.~\ref{sec-BB84}.
% by regarding the MU and Bob in MDIQKD as Bob in BB84.
In Sec.~\ref{simulation}, we give a numerical simulation on the proposed scheme, which is also compared to the original MDIQKD scheme and
the qubit-MDIQKD scheme of
Ref. \cite{Yin:QMDIQKD:2013}.
We also show the performance of BB84 with mismatched-basis statistics (our new scheme) and compare it with the original BB84.
%protocol given in Ref. \cite{Yin:QMDIQKD:2013}.
Finally, we conclude in Sec.~\ref{conclusion}.

\section{MDIQKD protocol with mismatched-basis statistics and main result} \label{main result}

\begin{figure}[hbt]
\centering
\resizebox{4in}{!}{\includegraphics{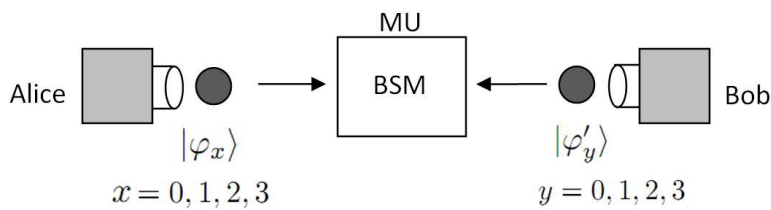}}
\caption{A schematic diagram for the MDIQKD protocol. BSM: Bell-state measurement, which is an untrusted device and may be controlled by Eve; $\ket{\varphi_0}$, $\ket{\varphi_1}$, $\ket{\varphi_2}$, $\ket{\varphi_3}$ ($\ket{\varphi'_0}$, $\ket{\varphi'_1}$, $\ket{\varphi'_2}$, $\ket{\varphi'_3}$) represent Alice (Bob)'s four encoding states.}
\label{Fig:MIsource:MIdiag}
\end{figure}

The protocol setting for MDIQKD with uncharacterized qubit sources is as follows.
Alice and Bob send their encoded qubits to
%Charlies (or Eve)
Eve
for BSM, as shown in Fig.~\ref{Fig:MIsource:MIdiag}.
When Alice (Bob) selects to output a state with index $x$ ($y$), her (his) encoding device emits a mixed qubit state $\rho_{\text{A},x}$ ($\rho_{\text{B},y}$) to Eve. Alice and Bob do not know what these states are.
For simplicity, we assume that the states are pure states $\rho_{\text{A},x}=\ket{\varphi_x}\bra{\varphi_x}$ and
$\rho_{\text{B},y}=\ket{\varphi'_y}\bra{\varphi'_y}$.
This is without loss of generality, and the mixed-state case automatically holds by using the same argument as in our qubit-MDIQKD analysis~\cite{Yin:QMDIQKD:2013}.
%The basic assumption is same as Ref. \cite{Yin:QMDIQKD:2013}:
We assume that
the initial joint state with Eve's system is
$
\rho_{\text{A},x} \otimes \rho_{\text{B},y} \otimes \rho_\text{E}
\:\:
\text{for all $x,y$}
$
where Eve's state $\rho_\text{E}$ is independent of $x$ and $y$.
The MU performs a BSM on the incoming states and announces whether the projection is successful to Alice and Bob.
The MU is required only to identity one Bell state (same as the original MDIQKD and unlike qubit-MDIQKD \cite{Yin:QMDIQKD:2013}).

We analyze its security using the entanglement distillation protocol (EDP) method \cite{Lo:QKDSecurity:1999,Shor:Preskill:2000}, which is widely used for security proofs of QKD. The essence of this method is to
regard our protocol as one that generates entangled pairs at the end. This means that we construct an equivalent EDP.  Then based on this EDP, we obtain the relation between the phase error rate and the bit error rate, the latter of which can be estimated in experiments.
Finally, the key generate rate can be calculated using this relation.
%convert our protocol into an equivalent EDP, and obtain the relation between the phase error rate and the bit error rate, which can be estimated in experiments.
We give the equivalent EDP version of our protocol as follows.

\begin{enumerate}
\item
Alice and Bob prepare $N$ pairs of entangled states,
\begin{equation} \label{MIs:Pure:iniState}
\begin{aligned}
\ket{\phi^+}_{AC} &=(\ket{0}_A|\varphi_0\rangle_C+\ket{1}_A|\varphi_1\rangle_C +|2\rangle_A|\varphi_2\rangle_C+|3\rangle_A|\varphi_3\rangle_C)/2, \\
\ket{\phi^+}_{BD} &=(\ket{0}_B|\varphi'_0\rangle_D+\ket{1}_B|\varphi'_1\rangle_D +|2\rangle_B|\varphi'_2\rangle_D+|3\rangle_B|\varphi'_3\rangle_D)/2, \\
\end{aligned}
\end{equation}
respectively. The subscripts $A$ and $B$ denote Alice's and Bob's classical raw key bits, respectively,
where we assign values $0$ and $1$ to basis $0$ and values $2$ and $3$ to basis $1$.
%, of which values $0$ and $1$ represent encodings in basis $0$, and values $2$ and $3$ represent encodings in basis $1$.
The states $|\varphi_x\rangle_C$ and $|\varphi'_x\rangle_D$ ($x=0,1,2,3$) are, respectively,  Alice's and Bob's uncharacterized encoding qubits to be sent to the MU. Alice and Bob are only know that $|\varphi_x\rangle_C$ and $|\varphi'_x\rangle_D$ are 2-dimensional states but do not know the details, since they do not trust
the accuracies of their encoding systems.
%the accuracy of their phase modulators.
%Here, we describe Alice's and Bob's encoding systems in an equivalent measurement-based way.
%It can be seen that
Essentially, Alice's and Bob's emitted states are determined by a measurement.
By measuring her half of the system, Alice collapses the system $C$ to one of $\ket{\varphi_i}_C$ with $i=0,1,2,3$ with equal probabilities, which is equivalent to Alice preparing the system $C$ in one of the four states with equal probabilities; similarly for Bob.
% The same argument holds for Bob.
%Hence, Eq.~\eqref{MIs:Pure:iniState} describes
%shows a mathematical equivalent description of
%Alice's and Bob's encoding systems.

\item
Alice and Bob send the states, labeled by $C$ and $D$ respectively, to Eve who announces her BSM result. There are two possible outcomes: BSM failure, or a successful measurement result in the Bell states
\begin{eqnarray}
|\phi^+\rangle_{CD} &=& (\ket{0}_C\ket{0}_D+\ket{1}_C\ket{1}_D)/\sqrt{2}. \label{MIs:Pure:BellState1}
\end{eqnarray}
For the first outcome Eve announces message $z=0$ to Alice and Bob, while message $z=1$ is announced for the second outcome. One must note that Eve might not honestly announce her measurement results, or might not even perform the above mentioned BSM. However, Eve must announce a message $z=0$ or $z=1$ to Alice and Bob for each trial.

\item
After receiving Eve's message, Alice and Bob perform bit sift: they discard their bits when Eve announces a BSM failure ($z=0$). Then, they project systems $A$ and $B$ in Eq.~\eqref{MIs:Pure:iniState} onto $\ket{0}\bra{0}+\ket{1}\bra{1}$ or $\ket{2}\bra{2}+\ket{3}\bra{3}$, which correspond to basis $0$ and basis $1$, respectively.
They perform basis sift next: when their systems collapse onto the same bases, by sacrificing some bits for error testing\footnote{An alternative way to do that is by performing error verification after error correction \cite{Fung:Finite:2010,MXF:Finite:2011}.}, they can deduce conditional probability distributions $p(z|x,y)$ where $z=0,1$ stands for Eve's announcements (failure or Eq.~\eqref{MIs:Pure:BellState1} respectively), and $x$ and $y$ ($x,y\in\{0,1,2,3\}$) represent the states of systems $A$ and $B$. When their systems collapse onto different bases, they deduce  similar probability distributions $p(z|x,y)$, but the raw key bits are discarded.

\item
Finally, Alice and Bob perform EDP on systems $A$ and $B$ and obtain maximally entangled Bell states $|\phi^{+\theta}\rangle_{AB}=(\ket{0}_A\ket{0}_B+e^{i\theta}\ket{1}_A\ket{1}_B)/\sqrt{2}$, where secret key bits can be extracted.
\end{enumerate}

The conditional probabilities of the measurement result by a lossless MU are listed in Table \ref{Tab:MIs:prob1}, from which one can see that there is a 75\% intrinsic loss for the original MDIQKD scheme when only one Bell state can be distinguished.

\begin{table}[htb]
\centering
\caption{List of conditional probabilities $p(z|x,y)$ for the case where Alice and Bob choose one of the four BB84 states with equal probabilities. Only cases where they choose the same basis are considered. No loss is considered.}
\label{Tab:MIs:prob1}
\begin{tabular}{c|cccc|cccc}
\backslashbox {$z$}{$x,y$}  & 0,0 & 0,1 & 1,0 & 1,1 & 2,2 & 2,3 & 3,2 & 3,3 \\
\hline
0 & 1/2 & 1 & 1 & 1/2 & 1/2 & 1 & 1 & 1/2 \\
1 & 1/2 & 0 & 0 & 1/2 & 1/2 & 0 & 0 & 1/2 \\
\hline \hline
\backslashbox {$z$}{$x,y$}  & 0,2 & 0,3 & 1,2 & 1,3 & 2,0 & 3,0 & 2,1 & 3,1 \\
\hline
0 & 1/4 & 1/4 & 1/4 & 1/4 & 1/4 & 1/4 & 1/4 & 1/4 \\
1 & 3/4 & 3/4 & 3/4 & 3/4 & 3/4 & 3/4 & 3/4 & 3/4 \\
\end{tabular}
\end{table}

Before introducing our security proof, let us see why statistics of mismatched basis cases can be used to generate secure key bits with a simple example. In the original MDIQKD protocol, Alice and Bob extract a secure key from the first half (matched-basis case) of the results shown in Table \ref{Tab:MIs:prob1}, and discard the second half (mismatched-basis case). Such postprocessing would fail when Alice and Bob do not trust the accuracy of their
qubit encoding systems.
%phase modulators (qubit states).
Consider the case when $|\varphi_0\rangle=|\varphi_2\rangle=|0\rangle$ and $|\varphi_1\rangle=|\varphi_3\rangle=|1\rangle$ for both Alice and Bob. It is obvious that Alice and Bob may still observe probabilities $p(z|x,y)$ with perfect correlations for the matched-basis case, but all key bits can be eavesdropped by Eve. On the other hand, the results from the mismatched-basis, the second half of Table \ref{Tab:MIs:prob1}, can be used to exclude this attack, in which Alice and Bob would find perfect correlations instead of random results.

\textbf{Main Result:} In above protocol, if Alice and Bob observe the probabilities $p(z|x,y)$, their final secret key bits in basis $0$ is given by $R=1-H(e_b)-H(e_p)$, in which $H(x)=-x \log x -(1-x) \log (1-x)$ is the Shannon's binary entropy function. The bit error rate in basis $0$ is given by
\begin{equation} \label{MIqubit:biterror}
\begin{aligned}
e_b=\frac{p(1|0,1)+p(1|1,0)}{p(1|0,0)+p(1|1,1)+p(1|0,1)+p(1|1,0)},
\end{aligned}
\end{equation}
and phase error rate is bounded by
\begin{equation}
\label{eqn-main-ep-eb-relation}
e_p\leqslant\varepsilon+e_b.
\end{equation}
The deviation $\varepsilon$ is defined as
%and calculated by the following ways.
\begin{equation}
\label{eqn-max-exp-result}
\begin{aligned}
\varepsilon \triangleq \max_{C,C'}f(C,C'),
\end{aligned}
\end{equation}
where the maximization takes over all non-negative real numbers $C_{30}$, $C_{31}$, $C'_{20}$ and $C'_{21}$ satisfying definite constraints to find the maximum value of function $f(C,C')$,
\begin{equation}
\label{eqn-max-exp}
\begin{aligned}
&f(C,C')=\\
&\begin{cases}
&\min\{\frac{\big(\sqrt{p(1|3,2)}+\sqrt{p(1|0,1)}C_{30}C'_{21}+\sqrt{p(1|1,0)}C_{31}C'_{20}+\sqrt{p(1|1,1)}\big|C_{30}C'_{20}-C_{31}C'_{21}\big|\big)^2}{ 2(p(1|0,0)+p(1|1,1)+p(1|0,1)+p(1|1,0))C^2_{30}C'^2_{20}},\\
&\frac{\big(\sqrt{p(1|3,2)}+\sqrt{p(1|0,1)}C_{30}C'_{21}+\sqrt{p(1|1,0)}C_{31}C'_{20}+\sqrt{p(1|0,0)}\big|C_{30}C'_{20}-C_{31}C'_{21}\big|\big)^2}{ 2(p(1|0,0)+p(1|1,1)+p(1|0,1)+p(1|1,0))C^2_{31}C'^2_{21}}\},\ \text{if}\ C_{30}C'_{20}\ne 0\ \text{and}\  C_{31}C'_{21}\ne 0\\
&\frac{\big(\sqrt{p(1|3,2)}+\sqrt{p(1|0,1)}C_{30}C'_{21}+\sqrt{p(1|1,0)}C_{31}C'_{20}+\sqrt{p(1|1,1)}\big|C_{30}C'_{20}-C_{31}C'_{21}\big|\big)^2}{ 2(p(1|0,0)+p(1|1,1)+p(1|0,1)+p(1|1,0))C^2_{30}C'^2_{20}},\ \text{if}\ C_{30}C'_{20}\ne 0\ \text{and}\ C_{31}C'_{21}= 0\\
&\frac{\big(\sqrt{p(1|3,2)}+\sqrt{p(1|0,1)}C_{30}C'_{21}+\sqrt{p(1|1,0)}C_{31}C'_{20}+\sqrt{p(1|0,0)}\big|C_{30}C'_{20}-C_{31}C'_{21}\big|\big)^2}{ 2(p(1|0,0)+p(1|1,1)+p(1|0,1)+p(1|1,0))C^2_{31}C'^2_{21}},\ \text{if}\ C_{30}C'_{20}= 0\ \text{and}\ C_{31}C'_{21}\ne 0\\
&1-e_b,\ \text{if}\  C_{30}C'_{20}= 0\ \text{and}\ C_{31}C'_{21}= 0, \\
\end{cases}
\end{aligned}
\end{equation}
where $\min\{a,b\}$ yields the smaller one of real numbers $a$ and $b$. And constraints for searching the maximum value are
\begin{equation}
\label{eqn-constraints1}
\begin{aligned}
&-2\sqrt{p(1|0,0)p(1|1,0)}C_{30}C_{31}\leqslant p(1|3,0)-p(1|0,0)C^2_{30}-p(1|1,0)C^2_{31}\leqslant 2\sqrt{p(1|0,0)p(1|1,0)}C_{30}C_{31}\\
&-2\sqrt{p(1|0,1)p(1|1,1)}C_{30}C_{31}\leqslant p(1|3,1)-p(1|0,1)C^2_{30}-p(1|1,1)C^2_{31}\leqslant 2\sqrt{p(1|0,1)p(1|1,1)}C_{30}C_{31}\\
&-2\sqrt{p(1|0,0)p(1|0,1)}C'_{20}C'_{21}\leqslant p(1|0,2)-p(1|0,0)C'^2_{20}-p(1|0,1)C'^2_{21}\leqslant 2\sqrt{p(1|0,0)p(1|0,1)}C'_{20}C'_{21}\\
&-2\sqrt{p(1|1,0)p(1|1,1)}C'_{20}C'_{21}\leqslant p(1|1,2)-p(1|1,0)C'^2_{20}-p(1|1,1)C'^2_{21}\leqslant 2\sqrt{p(1|1,0)p(1|1,1)}C'_{20}C'_{21}.\\
\end{aligned}
\end{equation}
We can see that the cross-basis statistics restrict the variables $C_{30}$, $C_{31}$, $C'_{20}$ and $C'_{21}$.
Thus, the phase error rate is obtained by numerical optimization.
The proof of this main result is detailed in appendix. Note that with the procedure introduced in the Sec. IV of Ref.~\cite{Yin:QMDIQKD:2013}, our main result applies to case that Alice's and Bob's encoding states are two-dimensional mixed states, although the proof given in appendix is based on two-dimensional pure states.

Above results are general results for arbitrary observed probabilities $p(z|x,y)$, but it seems a bit complicated. For ease of understanding, we simplify our results for typical MDIQKD implementations. Consider in a successful experiment for MDIQKD, one may observe that $p(1|00)=p(1|11)$, $p(1|01)=p(1|10)$ and $p(1|3,0)=p(1|3,1)=p(1|0,2)=p(1|1,2)=(p(1|00)+p(1|01))/2$. Under this case, we simplify the function $f(C,C')$ and the constraints as

\begin{equation}
\label{eqn-max-exp-s}
\begin{aligned}
&f(C,C')=\\
&\begin{cases}
&\frac{\big(\sqrt{e'_b}+\sqrt{e_b}(C_{30}C'_{21}+C_{31}C'_{20})+\sqrt{1-e_b}\big|C_{30}C'_{20}-C_{31}C'_{21}\big|\big)^2}{ 4max\{C^2_{30}C'^2_{20},{C^2_{31}C'^2_{21}\}}},\text{if}\ C_{30}C'_{20}\ne 0\ \text{and}\  C_{31}C'_{21}\ne 0\\
&\frac{\big(\sqrt{e'_b}+\sqrt{e_b}(C_{30}C'_{21}+C_{31}C'_{20})+\sqrt{1-e_b}\big|C_{30}C'_{20}-C_{31}C'_{21}\big|\big)^2}{ 4C^2_{30}C'^2_{20}},\ \text{if}\ C_{30}C'_{20}\ne 0\ \text{and}\ C_{31}C'_{21}= 0\\
&\frac{\big(\sqrt{e'_b}+\sqrt{e_b}(C_{30}C'_{21}+C_{31}C'_{20})+\sqrt{1-e_b}\big|C_{30}C'_{20}-C_{31}C'_{21}\big|\big)^2}{ 4C^2_{31}C'^2_{21}},\ \text{if}\ C_{30}C'_{20}= 0\ \text{and}\ C_{31}C'_{21}\ne 0\\
&1-e_b,\ \text{if}\  C_{30}C'_{20}= 0\ \text{and}\ C_{31}C'_{21}= 0, \\
\end{cases}
\end{aligned}
\end{equation}
where $e'_b=p(1|3,2)/(p(1|0,0)+p(1|0,1))$, and $max\{a,b\}$ yields the larger one of real numbers $a$ and $b$. And constraints for searching maximum value are simplified as
\begin{equation}
\label{eqn-constraints1-s}
\begin{aligned}
&-2\sqrt{e_b(1-e_b)}C_{30}C_{31}\leqslant \frac{1}{2}-(1-e_b)C^2_{30}-e_bC^2_{31}\leqslant 2\sqrt{e_b(1-e_b)}C_{30}C_{31}\\
&-2\sqrt{e_b(1-e_b)}C_{30}C_{31}\leqslant \frac{1}{2}-e_bC^2_{30}-(1-e_b)C^2_{31}\leqslant 2\sqrt{e_b(1-e_b)}C_{30}C_{31}\\
&-2\sqrt{e_b(1-e_b)}C'_{20}C'_{21}\leqslant \frac{1}{2}-(1-e_b)C'^2_{20}-e_bC'^2_{21}\leqslant 2\sqrt{e_b(1-e_b)}C'_{20}C'_{21}\\
&-2\sqrt{e_b(1-e_b)}C'_{20}C'_{21}\leqslant \frac{1}{2}-e_bC'^2_{20}-(1-e_b)C'^2_{21}\leqslant 2\sqrt{e_b(1-e_b)}C'_{20}C'_{21}.\\
\end{aligned}
\end{equation}

%Next, we give a simulation of our protocol.

\section{BB84 with uncharacterized sources and measurements}
\label{sec-BB84}

%$$\frac{|00\rangle+|11\rangle}{\sqrt{2}}$$
Our MDIQKD security analysis can be directly applied to the BB84 protocol with the following conditions:
\begin{itemize}
\item
Alice prepares one of four uncharacterized qubit states;
\item
Bob receives a qubit state from the channel;
\item
and Bob's measurement is one of two uncharacterized qubit von Neumann (projective) measurements corresponding to the two BB84 bases.
\end{itemize}
%and Bob measures them with uncharacterized projection measurements.
%a complete POVM measurement with two outcomes.
This means that
%We assume that
the measurement for each basis is a projection onto two orthogonal states.
%Also, the detection inefficiency is assumed to be the same for the two measurements and is absorbed into the channel loss.
%In BB84,
Let Bob's qubit measurement for basis $0$ be a projection onto
$\{|\bar\varphi'_0\rangle,|\bar\varphi'_1\rangle\}$
and
for basis $1$ be
$\{|\bar\varphi'_2\rangle,|\bar\varphi'_3\rangle\}$
where $\langle \bar\varphi'_0|\bar\varphi'_1\rangle=\langle \bar\varphi'_2|\bar\varphi'_3\rangle=0$ and
$
\ket{\bar\varphi'_0}\bra{\bar\varphi'_0}+
\ket{\bar\varphi'_1}\bra{\bar\varphi'_1}
=
\ket{\bar\varphi'_2}\bra{\bar\varphi'_2}+
\ket{\bar\varphi'_3}\bra{\bar\varphi'_3}
=
I$.

\begin{figure}[hbt]
\centering
%\resizebox{4in}{!}
{\includegraphics[width=\linewidth]{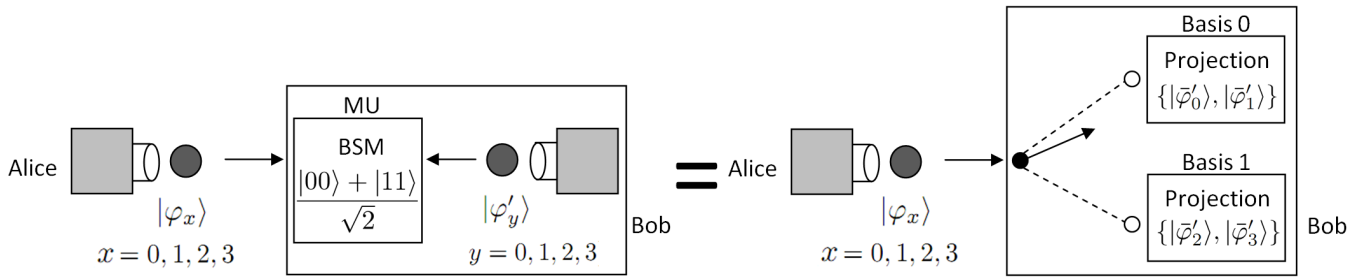}}
\caption{The security proof of MDIQKD (left) can be applied to BB84 (right) when the MU is merged into Bob.  We impose that the MU honestly announces the $|\phi^+\rangle=(|00\rangle+|11\rangle)/\sqrt{2}$ outcome when it performs the BSM.
The two von Neumann (projective) measurements of Bob on the right correspond to the two BB84 bases and each projects onto two orthogonal qubit states.
The relationship between the states in MDIQKD and BB84 is that
$\ket{\varphi'_y}$ is the complex conjugate of $\ket{\bar\varphi'_y}$ for $y=0,1,2,3$.
}
\label{Fig:BB84:equiv}
\end{figure}

The main idea is to merge the MU and Bob in the MDIQKD setting to become Bob in the BB84 setting (see Fig.~\ref{Fig:BB84:equiv}).
%Then, projecting the Bell state measurement of the MU onto Bob's output state
%%in the MDIQKD setting becomes
%gives
%a qubit measurement to be performed on Alice's state.
%We detail the idea as follows.
%M_{\text{loss},0}
%$\langle \bar\varphi'_2|\bar\varphi'_3\rangle=0$, and
%Bob's qubit measurement for basis $0$ is a POVM with elements $\{M_0,M_1\}$ and
%for basis $1$ is $\{M_2,M_3\}$, where $M_0+M_1=I$ and $M_2+M_3=I$.
%Note that the detection inefficiency is assumed to be basis independent and is absorbed into the channel loss.
%Bob's qubit measurement for basis $0$ is a POVM with elements $\{L^{(0)},M_0,M_1\}$ and
%for basis $1$ is $\{L^{(1)},M_2,M_3\}$, where $L^{(0)}+M_0+M_1=I$ and $L^{(1)}+M_2+M_3=I$.
%Here, $L^{(b)}, b=0,1$ represent losses due to the channel or the detectors.
%We first assume that $M_i=\alpha_i |\varphi'_i\rangle \langle\varphi'_i|$, $i=0,1,2,3$ are projections onto arbitrary qubit states with efficiencies given by $\alpha_i$.
%We first assume that $M_y= |\varphi'_y\rangle \langle\varphi'_y|$, $y=0,1,2,3$ are projections onto arbitrary pure qubit states.
%
%
In the BB84 picture,
Alice emits a qubit state which is processed by Eve and is received by Bob as $\rho_x$.
Bob chooses basis 0 or 1 with equal probabilities to measure it.
The probability of obtaining
$M_y= |\bar\varphi'_y\rangle \langle\bar\varphi'_y|$, $y=0,1,2,3$ conditional on a chosen basis is
%$M_y$ is
%$q(y|x)=\tr (\rho_x M_y)$.
$\tr (\rho_x M_y)$.
% where $\rho_x$ is the qubit state coming from Alice via the channel.
%where $q_0+q_1=q_2+q_3=1$.

Alternatively, Bob may do the following to perform effectively the same measurement.
Bob prepares one of the four states $|\varphi'_y\rangle$, $y=0,1,2,3$, with equal probabilities and makes a BSM on
$\rho_x \otimes \ket{\varphi'_y}\bra{\varphi'_y}$.
If the BSM produces the projection outcome of $|\phi^+\rangle=(|00\rangle+|11\rangle)/\sqrt{2}$, then this is equivalent to measuring $\rho_x$ with $M_y$.
The key to this argument is to notice that
$$
(\bra{\varphi_x}_A \bra{\varphi'_y}_B) (|00\rangle+|11\rangle)_{AB}= \braket{\varphi_x}{\bar\varphi'_y}_A
$$
for any $\ket{\varphi_x}$ and $\ket{\varphi'_y}$ where
$\ket{\bar\varphi'_y}$ is the complex conjugate of $\ket{\varphi'_y}$.
This means that
the probability of obtaining a $\ket{\phi^+}$ projection by the BSM is
\begin{align*}
p(1|x,y)&=
\tr \left(\ket{\phi^+}\bra{\phi^+} (\rho_x\otimes \ket{\varphi'_y}\bra{\varphi'_y})\right)
\\
&=
\frac{1}{2}\tr (\rho_x M_y).
\end{align*}
The term $p(1|x,y)$ corresponds to a hypothetical MDIQKD setting where the MU identifies $\ket{\phi^+}$ and the last term corresponds to the BB84 setting.
%This shows that the MU probability in the MDIQKD setting is equal to half of Bob's measurement probability in the BB84 setting.
%\footnote{%
%Though not necessary, we could assign the probability of Bob choosing either basis in BB84 to be half and then the equivalence would hold in a strict sense.}.
Thus, we can regard the BB84 setting where we drop half of the measurement outcomes as an MDIQKD setting where the measurement outcomes are kept when the MU gets $\ket{\phi^+}$.
To make this equivalence rigorous,
we also need the freedom to choose the distribution for selecting $y$ in the MDIQKD setting to match the occurrences of $y$ in the BB84 setting.
We have this freedom in the MDIQKD setting
%which is valid
because the phase error bound in our security proof is independent of this distribution.
%
%with MU probabilities given by the above formula (simply multiply by half with the BB84 measurement probabilities).
%Thus, from the BB84 measurement probabilities, we can construct a hypothetical MDIQKD situation with MU probabilities given by the above formula (simply multiply by half).
Since the MDIQKD situation
is a restricted case of the general one considered in our proof (because we consider a specific MU that identifies $\ket{\phi^+}$ honestly and Eve is allowed to intervene with Alice's state only), it is covered by our proof,
%Our MDIQKD proof does not depend on the distribution over $y$.
%depends on t
which means that the equivalent BB84 setting where half of the outcomes are dropped is also covered by our proof.
To get a relationship between bit and phase error rates for the actual BB84 protocol, we simply obtain the measurement probabilities $\tr (\rho_x M_y)$ in the experiment and
use them as $p(1|x,y)$ in the formulas \eqref{eqn-main-ep-eb-relation}-\eqref{eqn-constraints1}, since the factor of half will be canceled out anyway.
Note that in the actual BB84 protocol, there is no need to drop half of the outcomes because both the dropped half and the retained half have the same statistics anyway.
We could have kept the dropped half and would have obtained the exact same phase error bound.
Thus, the two halves can be used together to generate a secret key.

%Now, we allow the POVM elements $M_y=\sum_k \alpha_{y,k}|\varphi'_{y,k}\rangle \langle\varphi'_{y,k}|$ to be of arbitrary rank.
%This correspond to Bob sending a mix state $\sum_k \alpha_{y,k}\ket{\bar \varphi'_{y,k}}\bra{\bar \varphi'_{y,k}}$ in the equivalent MDIQKD picture.
%Arguing similarly as above and since our MDIQKD proof applies to mix states as well, we can reuse the proof in the BB84 setting by treating the measurement probabilities $\tr (\rho_x M_y)$ as $p(1|x,y)$.

%multiply by half to obtain $p(1|x,y)$ which are
%We projecting this onto Alice's state $|\varphi_x\rangle$ and Bob's state $|\varphi'_y\rangle$ gives the conditional probability $p(1|xy)=\langle \varphi'_y| \varphi_x \rangle$.
%On the other hand, the conditional probability of getting the $M_y$ result is
%$\tr (|\varphi_x\rangle \langle\varphi_x| M_y)$

\section{Simulation}
\label{simulation}

We first consider the case that Alice and Bob use ideal BB84 senders (not trusted by Alice and Bob though) and an ideal Bell-state MU to perform
%QMDIQKD
our new protocol
with noiseless channel and no Eve's attack, but with photon absorption taken into consideration. One must observe that $e_b=0$, $p(1|3,0)=p(1|3,1)=p(1|0,2)=p(1|1,2)=p(1|0,0)/2=p(1|1,1)/2$, then with
constraints \eqref{eqn-constraints1}, we deduce that $C_{30}=C_{31}=C'_{20}=C'_{21}=1/\sqrt{2}$.
Then, through \eqref{eqn-max-exp}, it is easy to verify that $\varepsilon=0$ and thus $e_p=0$. Hence, MDIQKD can be secure in this situation.

In general, an analytical expression for the key rate is hard to obtain when channel errors are presented. One can measure values of $p(z|x,y)$ from experiments directly and obtain the key rate (such value needed in postprocessing) via numerical methods. For comparison, we assume the errors appeared in the state preparation and measurement are absorbed into the channel, which can be controlled by Eve. In the simulation, four single photon detectors (SPDs) with the dark counting rate of $d$ are used in the MU. Denote $\eta$ as the total transmission efficiency of the channel from Alice (Bob) to the MU in the middle of channel and thus $\eta$ is also the probability that a single photon from Alice or Bob can trigger a SPD of the MU.

Consequently, if Alice and Bob both emit qubit $|0\rangle$ or $|1\rangle$, the MU will announce message $1$ with probability $p(1|0,0)=p(1|1,1)=\eta^2(1-d)^2/2+2\eta(1-\eta)d(1-d)^2+2(1-\eta)^2d^2(1-d)^2$, in which the first item corresponds to the case that the projection of the incoming photons into $|\phi^+\rangle_{CD}$ is successful: the two photons trigger the two SPDs and the remaining two SPDs do not give dark clicks. The second item accounts for the case that only one photon triggers one SPD but a dark count occurs in one relevant SPD, and the last item represents the case that two photons are absorbed by the channel but two dark counts occur in two relevant SPDs. And $p(0|0,0)=p(0|1,1)=1-p(1|0,0)$ also holds.

\begin{figure}[!t]
%\center{!}
{\includegraphics[width=.5\columnwidth]{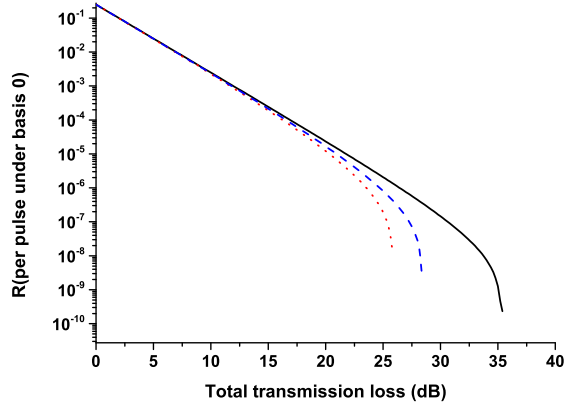}}
\caption{(Color online) Secure-key rate $R$ (unit: per pulse of basis $0$) vs total transmission loss (dB) of channel from Alice or Bob to MU when single photon sources are equipped: we set $d=10^{-5}$ per pulse.
The solid line represents MDIQKD with perfect trustworthy BB84 senders' devices,  MU without optics misalignment, which can distinguish one Bell state $|\phi^+\rangle_{CD}$, and no Eve's attack; the dashed line is for the protocol proposed here, in which perfect BB84 senders' devices (but not trusted by Alice and Bob), MU without optics misalignment, which can distinguish one Bell states $|\phi^+\rangle_{CD}$, and no Eve's attack;
the dotted line is for the qubit-MDIQKD protocol proposed in Ref. \cite{Yin:QMDIQKD:2013}, in which perfect BB84 senders' devices (but not trusted by Alice and Bob), MU without optics misalignment, which can distinguish two Bell states $|\phi^+\rangle_{CD}$ and $|\psi^+\rangle_{CD}$, and no Eve's attack.
\label{fig-1}}
\end{figure}

\begin{figure}[!t]
%\center{!}
{\includegraphics[width=.5\columnwidth]{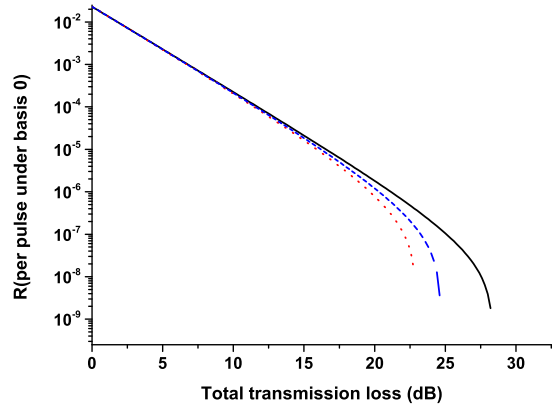}}
\caption{(Color online) Secure-key rate $R$ (unit: per pulse of basis $0$) vs total transmission loss (dB) of channel from Alice or Bob to MU when practical coherent sources are equipped: we set $d=10^{-5}$ per pulse.
The solid line represents MDIQKD with trustworthy coherent source with mean photon number $\mu=0.5$, MU without optics misalignment, which can distinguish one Bell state $|\phi^+\rangle_{CD}$, and no Eve's attack; the dashed line is for the protocol proposed here, in which coherent source with mean photon number $\mu=0.5$ (its encoding is not trusted by Alice and Bob but its photon statistics is trustworthy), MU without optics misalignment, which can distinguish one Bell states $|\phi^+\rangle_{CD}$, and no Eve's attack;
the dotted line is for the qubit-MDIQKD protocol proposed in Ref. \cite{Yin:QMDIQKD:2013}, in which coherent source with mean photon number $\mu=0.5$ (its encoding is not trusted by Alice and Bob but its photon statistics is trustworthy), MU without optics misalignment, which can distinguish two Bell states $|\phi^+\rangle_{CD}$ and $|\psi^+\rangle_{CD}$, and no Eve's attack. Infinite decoy states are employed here.
\label{fig-2}}
\end{figure}

 By the similar considerations, we set $p(1|1,0)=p(1|0,1)=p(1|3,2)=2(1-\eta)^2d^2(1-d)^2+2\eta(1-\eta)d(1-d)^2$, and $p(0|1,0)=p(0|0,1)=1-p(1|1,0)$, $p(1|3,0)=p(1|3,1)$ and $p(1|0,2)=p(1|1,2)$. Note that the formulas for the simulation are relevant to the existing literature (cf., Eq.(7) in \cite{MXF:MIFluc:2012}). The secure-key rate (unit: per pulse under basis $0$) versus total transmission loss of channel from Alice or Bob to the MU is given by Fig.~\ref{fig-1} and Fig.~\ref{fig-2}.

In Fig.~\ref{fig-1}, the secure-key rate for original MDIQKD is given by the solid line, in which we assume that Alice and Bob know that their encoding states are perfect, the MU can only distinguish Bell state $|\phi^+\rangle_{CD}$, and Eve is passive.
The secure-key rate for our protocol is given by the dashed line, in which
%Conversely, still with these
we have ideal BB84 senders (but we do not trust them now), MU can distinguish one Bell states $|\phi^+\rangle_{CD}$, and Eve is passive. In Fig.~\ref{fig-2}, we consider that practical coherent sources are used by Alice and Bob, and infinite decoy states\cite{Hwang:Decoy:2003,Lo:Decoy:2005,Wang:Decoy:2005} are employed.
%the key bits rate for QMDIQKD is given by dashed line in Fig.1 when Eve is passive.

\begin{figure}[!t]
%\center{!}
{\includegraphics[width=.5\columnwidth]{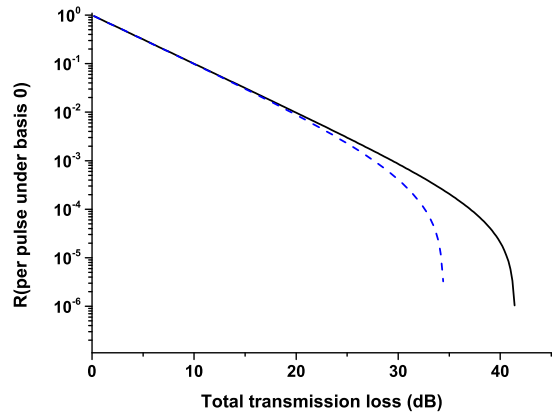}}
\caption{(Color online) Secure-key rate $R$ (unit: per pulse of basis $0$) vs total transmission loss (dB) of channel from Alice to Bob: we set $d=10^{-5}$ per pulse.
The solid line represents original BB84 protocol with perfect trustworthy BB84 senders' devices and measurement device; the dashed line is for the protocol proposed here, in which perfect BB84 senders' devices (but not trusted by Alice and Bob) are equipped while Bob's measurement devices are also uncharacterized 2-dimensional projections.
\label{fig-3}}
\end{figure}

 Additionally, a numerical simulation for BB84 with uncharacterized qubit sources and measurements is given in Fig.~\ref{fig-3}. In this simulation, we use the same parameters as the simulation for MDIQKD.

\begin{figure}[!t]
%\center{!}
{\includegraphics[width=.5\columnwidth]{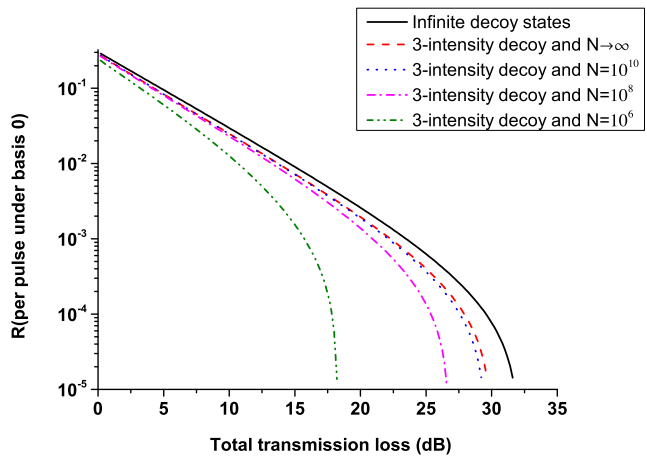}}
\caption{(Color online) Secure-key rate $R$ (unit: per pulse of basis $0$) vs total transmission loss (dB) of channel from Alice to Bob: we set $d=10^{-5}$ per pulse.
The solid line represents the proposed BB84 protocol with infinite decoy states; Other lines are all for the three-decoy-state protocol (i.e., vacuum states + weak decoy states +signal states) and the mean photon number for decoy states and signal states are 0.1 and 0.5 respectively. the dashed line, dotted line, dashed-dotted line and dashed-dotted-dotted line are for $N=\infty,10^{10},10^8,10^6$ respectively.
\label{fig-4}}
\end{figure}

In practice, only a finite number of decoy states are used.
% To perform our protocol in practical, infinite decoy states method is not enough.
 We simulate our proposed BB84 case with three decoy states  (i.e., vacuum state + weak decoy state + signal state) and with statistical fluctuations in Fig.~\ref{fig-4}. We assume that the mean photon number of the Poisson-distributed weak decoy state and signal state are 0.1 and 0.5 respectively. The pulse number of each encoding states is set to $N$ and 5-times standard derivation is considered.

\begin{figure}[!t]
%\center{!}
{\includegraphics[width=.5\columnwidth]{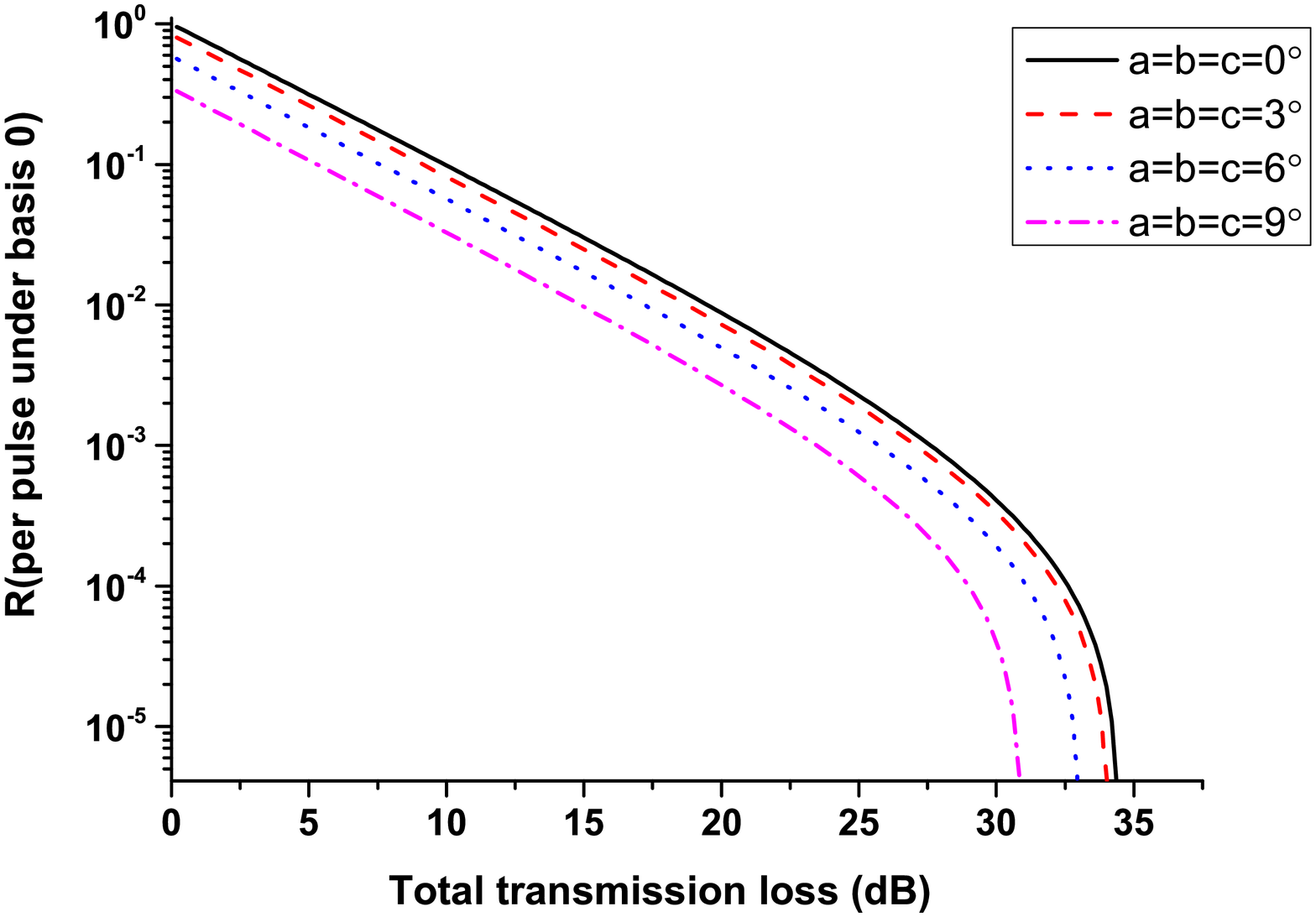}}
\caption{(Color online) Secure-key rate $R$ (unit: per pulse of basis $0$) vs total transmission loss (dB) of channel from Alice to Bob: we set $d=10^{-5}$ per pulse.
The solid line represents the proposed BB84 protocol without encoding misalignment; the dashed line is for the case that $a=b=c=3^\circ$, the dotted line is for the case that $a=b=c=6^\circ$, and the dashed-dotted line is for the case that $a=b=c=9^\circ$.
\label{fig-5}}
\end{figure}

The above simulations consider the cases with no encoding misalignments. However, encoding misalignments are inevitable in QKD systems and the power of our method is that we do not need to characterize misalignment errors. For simplicity, we give a numerical simulation for BB84 protocol with the following typical encoding misalignments. We assume that Alice's encoding system prepares quantum states $\ket{\varphi_0}=\ket{0}$, $\ket{\varphi_1}=\sin a\ket{0}+\cos a\ket{1}$, $\ket{\varphi_2}=\cos (\pi/4+b)\ket{0}+\sin (\pi/4+b)\ket{1}$, and $\ket{\varphi_3}=\sin (\pi/4+c)\ket{0}-\cos (\pi/4+c)\ket{1}$ for inputting $x=0,1,2,3$. Here, the degrees of angles $a$, $b$ and $c$ are the encoding misalignments. Without loss of generality, we assume that Bob's measurements are ideal BB84 measurements without misalignment. Consequently, we have that $p(1|0,0)=\eta (1-p_d)+(1-\eta)p_d(1-p_d)$, $p(1|1,1)=(1-p_d)\eta \cos^2 a+(1-\eta)p_d(1-p_d)$, $p(1|0,1)=(1-\eta)p_d(1-p_d)$, $p(1|1,0)=(1-p_d)\eta\sin^2 a+(1-\eta)p_d(1-p_d)$, $p(1|3,2)=(1-p_d)\eta(\sin(\pi/4+c)-\cos(\pi/4+c))^2/2+(1-\eta)p_d(1-p_d)$, $p(1|3,0)=(1-p_d)\eta\sin^2(\pi/4+c)+(1-\eta)p_d(1-p_d)$,
 $p(1|3,1)=(1-p_d)\eta\cos^2(\pi/4+c)+(1-\eta)p_d(1-p_d)$, $p(1|0,2)=(1-p_d)\eta/2+(1-\eta)p_d(1-p_d)$, and $p(1|1,2)=(1-p_d)\eta(\sin a+\cos a)^2/2+(1-\eta)p_d(1-p_d)$. The key rates with different misalignments and channel loss are illustrated in Fig.~\ref{fig-5}, from which we can see that with the help of mismatched-basis statistics the secure key rate only degrades with misalignment errors $a,b,c$ slightly. And, within a reasonable misalignment range, our protocol is still very practical.

\section{Conclusion} \label{conclusion}

In this paper, we propose that the statistics for the bits from mismatched bases can relax the assumptions on the encoding devices in original MDIQKD and BB84 protocols.
Our method does not need any modification of the original MDIQKD or BB84 protocol, except that Alice and Bob should obtain some statistics of bits from mismatched bases. Alice and Bob do not need to guarantee the accuracies of their encoding systems except that they need to be sure that the encoding states are two-dimensional. Note that our assumptions satisfy many practical MDIQKD and BB84 systems. For example, in phase encoding systems, it is reasonable to assume that encoding states are in two-dimensional space while the accuracies of the phase modulators may be questionable. The simulation results show that with decoy states, our scheme can distribute secure key bits over long distances (over 100km).
Our main proof is for the MDIQKD protocol, but we show that it can
%Finally, we show how our proof for MDIQKD can
be used for the BB84 protocol with uncharacterized sources and measurements by considering the MU to be part of Bob.

\section*{Acknowledgments}
This work was supported by the National Basic Research Program of China (Grants No. 2011CBA00200 and No. 2011CB921200), National Natural Science Foundation of China (Grants No. 61101137, No. 61201239 and No. 61205118). X.~M.~gratefully acknowledges the financial support from the National Basic Research Program of China Grants No.~2011CBA00300 and No.~2011CBA00301; and the 1000 Youth Fellowship program in China. C.-H.~F.~F.~gratefully acknowledges the financial support of RGC Grant No.~700712P from the HKSAR Government.

%\section{appendix}
\appendix
\section{Proof of main result}
We prove our main result here.
Following a similar argument used in Ref.~\cite{Shor:Preskill:2000}, an EDP has been given in section II.
Given that the initial states of Alice's, Bob's and Eve's ancillas are separable,
the most general collective attack by Eve can be represented by a unitary transformation as follows:
%we express Eve's collective attack by a unitary transformation:
\begin{equation}
\label{eqn-model1}
\begin{aligned}
&U_{Eve}|\varphi_{x}\rangle_{C}|\varphi'_{y}\rangle_{D}
|e\rangle_{Ea}\ket{0}_M=\sqrt{p(0|x,y)}|\Gamma_{xy0}\rangle_E\ket{0}_M +\sqrt{p(1|x,y)}|\Gamma_{xy1}\rangle_E\ket{1}_M,
\end{aligned}
\end{equation}
where $x,y=\{0,1,2,3\}$, $|e\rangle_{Ea}$ is Eve's arbitrary ancilla, $\ket{0}_M$ is the message which will be sent to Alice and Bob, and $|\Gamma_{xy0}\rangle_{E}$ and $|\Gamma_{xy1}\rangle_{E}$ are all normalized Eve's arbitrary quantum states for Eve's ancilla and photons $C$, $D$.
Here,
%we emphasize
we remark
that our collective attack modeled by Eq.~\eqref{eqn-model1}
includes both
basis-independent attack and basis-dependent attack \cite{GLLP:2004}.
%has taken the basis-independent attack and basis-dependent attack \cite{GLLP:2004} into account.
In a basis-independent (-depedent) attack,
the density matrix of the states emitted by Alice and Bob for basis $0$ is the same as (different from) that for basis $1$.
%transmitted in the channel when Alice and Bob's bases choices are both $0$ is the same as the case of basis $1$.
%The basis-independent attack is a situation where the density matrix of the states transmitted in the channel when Alice and Bob's bases choices are both $0$ is the same as the case of basis $1$. The basis-dependent attack is a situation where the two density matrices are different.
%The basis-dependent attack is a situation that the initial density matrix of encoding states when Alice and Bob's bases choices are both $0$ is different from the initial density matrix of encoding states when Alice and Bob's bases choices are both $1$.
The basis dependence can be measured by fidelity. In general, for basis-dependent attacks, Eve may obtain basis information
by some measurements and adopt different operations accordingly.
%through certain measurements and then adopt different operations accordingly.
Any measurement that Eve may utilize to learn the basis and the followup operations can be seen as part of an extended unitary transformation on
%the encoding states and the necessary ancilla. Thus in this view, Eve's overall operation in the $0$ and $1$ bases can be described as a unitary transformation on
Alice's and Bob's encoding states and her ancilla given by Eq.~\eqref{eqn-model1}.

We again assume that Alice and Bob do not know the details of
$|\varphi_x\rangle_C$ and $|\varphi'_y\rangle_D$ $(x,y=0,1,2,3)$.

Recall that $|\varphi_x\rangle_C$ and $|\varphi'_y\rangle_D$
%$(x,y=0,1,2,3)$
are both in the two-dimensional Hilbert space and they are disjoint ($|\varphi_{xy}\rangle_{CD}=|\varphi_x\rangle_C|\varphi'_y\rangle_D$)
, we may arbitrarily assign a phase to each of them. Thus

\begin{equation}
\label{eqn-varphi23}
\begin{aligned}
|\varphi_2\rangle_C&=C_{20}|\varphi_0\rangle_C+C_{21}e^{i\theta_2}|\varphi_1\rangle_C \\
|\varphi_3\rangle_C&=C_{30}|\varphi_0\rangle_C+C_{31}e^{i\theta_3}|\varphi_1\rangle_C \\
|\varphi'_2\rangle_D&=C'_{20}|\varphi'_0\rangle_D+C'_{21}e^{i\theta'_2}|\varphi'_1\rangle_D \\
%\text{ and }\\
|\varphi'_3\rangle_D&=C'_{30}|\varphi'_0\rangle_D+C'_{31}e^{i\theta'_3}|\varphi'_1\rangle_D
\end{aligned}
\end{equation}
must hold for some non-negative real numbers $C_{xy}$ and $C'_{xy}$.

To begin our analysis, without loss of generality, we can assume in Eq.~\eqref{eqn-model1} that $|\Gamma_{xyz}\rangle_E=\sum_n\gamma_{xyzn}|n\rangle_E$, in which $|n\rangle_E$ are a set of normalized orthogonal bases of Eve's states, and complex number $\gamma_{xyzn}=_E\langle n|\Gamma_{xyz}\rangle_E$, satisfying $\sum_n\big|\gamma_{xyzn}\big|^2=1$. Thus, the density matrix for the case that Alice and Bob both select basis $0$ is
\begin{equation}
\label{eqn-rhoAB-1}
\begin{aligned}
\rho=&\frac{1}{p(1|0,0)+p(1|1,1)+p(1|0,1)+p(1|1,0)}\cdot \\
&\sum_nP\{\sqrt{p(1|0,0)}\gamma_{001n}\ket{0}_A\ket{0}_B+\sqrt{p(1|1,1)}\gamma_{111n}\ket{1}_A\ket{1}_B+\sqrt{p(1|0,1)}\gamma_{011n}\ket{0}_A\ket{1}_B+\sqrt{p(1|1,0)}\gamma_{101n}\ket{1}_A\ket{0}_B\}\\
&=\frac{\sum_nP\{\sqrt{p(1|0,0)}\gamma_{001n}\ket{0}_A\ket{0}_B+\sqrt{p(1|1,1)}\gamma_{111n}\ket{1}_A\ket{1}_B+\sqrt{p(1|0,1)}\gamma_{011n}\ket{0}_A\ket{1}_B+\sqrt{p(1|1,0)}\gamma_{101n}\ket{1}_A\ket{0}_B\}}{p(1|0,0)+p(1|1,1)+p(1|0,1)+p(1|1,0)},
\end{aligned}
\end{equation}
in which, $P\{|x\rangle\}=|x\rangle\langle x|$. The aim of this EDP is to obtain perfect Bell states $|\phi^{+\alpha}\rangle_{AB}=(\ket{0}_A\ket{0}_B+e^{i(\alpha_A+\alpha_B)}\ket{1}_A\ket{1}_B)/\sqrt{2}$. Accordingly, we can define the bit error rate $e_{b}$ and phase error rate $e_{p}$ under basis $0$:

%\begin{equation}
%\begin{aligned}
\begin{eqnarray}
e_b&=& _A\langle 0|_B\langle 1|\rho\ket{1}_B\ket{0}_A+_A\langle 1|_B\langle 0|\rho\ket{0}_B\ket{1}_A=\frac{p(1|0,1)+p(1|1,0)}{p(1|0,0)+p(1|1,1)+p(1|0,1)+p(1|1,0)}\\
e_p&=& _{AB}\langle \phi^{-\alpha}|\rho|\phi^{-\alpha}\rangle_{AB}+_{AB}\langle \psi^{-\alpha}|\rho|\psi^{-\alpha}\rangle_{AB}\nonumber
\label{eqn-eb-ep}
\\
&=&\frac{\sum_n\big|\sqrt{p(1|0,0)}\gamma_{001n}-e^{-i(\alpha_A+\alpha_B)}\sqrt{p(1|1,1)}\gamma_{111n}\big|^2+\sum_n\big|\sqrt{p(1|0,1)}\gamma_{011n}-e^{-i(\alpha_A-\alpha_B)}\sqrt{p(1|1,0)}\gamma_{101n}\big|^2}{2(p(1|0,0)+p(1|1,1)+p(1|0,1)+p(1|1,0))}
\nonumber
\\
&\leqslant &\frac{\sum_n\big|\sqrt{p(1|0,0)}\gamma_{001n}-e^{-i(\alpha_A+\alpha_B)}\sqrt{p(1|1,1)}\gamma_{111n}\big|^2}{2(p(1|0,0)+p(1|1,1)+p(1|0,1)+p(1|1,0))}+e_b
\end{eqnarray}
%\end{aligned}
%\end{equation}
in which, $|\phi^{-\alpha}\rangle_{AB}=(\ket{0}_A\ket{0}_B-e^{i(\alpha_A+\alpha_B)}\ket{1}_A\ket{1}_B)/\sqrt{2}$, and $|\psi^{-\alpha}\rangle_{AB}=(\ket{0}_A\ket{1}_B-e^{i(\alpha_A-\alpha_B)}\ket{1}_A\ket{0}_B)/\sqrt{2}$. The goal is to upper-bound $e_p$ effectively. Before proceeding, we remark that we just focus on the $e_p$ and final key bits rate for basis $0$ for simplicity, and thus we do not need to calculate the density matrix for basis $1$. But this basis $0$'s $e_p$ must be related to some probabilities of basis $1$, such as $p(1|3,2)$. Now we begin to detail how to obtain an upper bound of $e_p$.

We substitute the relations \eqref{eqn-varphi23} into Eq.~\eqref{eqn-model1} to obtain the following constraints:
\begin{equation}
\label{eqn-constraints1-A}
\begin{aligned}
&C_{x0}C'_{y0}\sqrt{p(z|0,0)}|\Gamma_{00z}\rangle_E
+C_{x0}C'_{y1}\sqrt{p(z|0,1)}e^{i\theta'_y}|\Gamma_{01z}\rangle_E\\
&+C_{x1}C'_{y0}\sqrt{p(z|1,0)}e^{i\theta_x}|\Gamma_{10z}\rangle_E
+C_{x1}C'_{y1}\sqrt{p(z|1,1)}e^{i(\theta_x+\theta'_y)}|\Gamma_{11z}\rangle_E\\&=\sqrt{p(z|x,y)}|\Gamma_{xyz}\rangle_E,
\end{aligned}
\end{equation}
in which
$x,y=\{2,3\}$
%$x,y=\{0,1,2,3\}$ but $x=y=0$ is excluded,
and $z=\{0,1\}$. Considering that $|\Gamma_{xyz}\rangle_E$ can be spanned by a set of basis $|n\rangle_E$ and with Eq.~\eqref{eqn-constraints1-A}, we obtain
\begin{equation}
\label{eqn-constraints2-A}
\begin{aligned}
&\sum_n\big|C_{30}C'_{20}\sqrt{p(1|0,0)}\gamma_{001n}+C_{31}C'_{21}\sqrt{p(1|1,1)}e^{i(\theta_3+\theta'_2)}\gamma_{111n}\big|^2\\
&\leqslant
(\sqrt{p(1|3,2)}+\sqrt{p(1|0,1)}C_{30}C'_{21}+\sqrt{p(1|1,0)}C_{31}C'_{20})^2.\\
\end{aligned}
\end{equation}
By observing the left-hand side of \eqref{eqn-constraints2-A} and with the help of triangle inequality
and Cauchy-Schwarz inequality, we have:

\begin{equation}
\begin{aligned}
&\sum_n\big|C_{30}C'_{20}\sqrt{p(1|0,0)}\gamma_{001n}+C_{31}C'_{21}\sqrt{p(1|1,1)}e^{i(\theta_3+\theta'_2)}\gamma_{111n}\big|^2\\
&\geqslant\sum_n (C_{30}C'_{20}\big|\sqrt{p(1|0,0)}\gamma_{001n}+\sqrt{p(1|1,1)}e^{i(\theta_3+\theta'_2)}\gamma_{111n}\big|-\big|C_{30}C'_{20}-C_{31}C'_{21}\big|\sqrt{p(1|1,1)}\big|\gamma_{111n}\big|)^2\\
&=C^2_{30}C'^2_{20}\sum_n\big|\sqrt{p(1|0,0)}\gamma_{001n}+\sqrt{p(1|1,1)}e^{i(\theta_3+\theta'_2)}\gamma_{111n}\big|^2\\
&+(C_{30}C'_{20}-C_{31}C'_{21})^2p(1|1,1)-2C_{30}C'_{20}\big|C_{30}C'_{20}-C_{31}C'_{21}\big|\sqrt{p(1|1,1)}\sum_n\big|\gamma_{001n}+e^{i(\theta_3+\theta'_2)}\gamma_{111n}\big|\big|\gamma_{111n}\big|\\
&\geqslant C^2_{30}C'^2_{20}\sum_n\big|\sqrt{p(1|0,0)}\gamma_{001n}+\sqrt{p(1|1,1)}e^{i(\theta_3+\theta'_2)}\gamma_{111n}\big|^2\\
&+(C_{30}C'_{20}-C_{31}C'_{21})^2p(1|1,1)-2C_{30}C'_{20}\big|C_{30}C'_{20}-C_{31}C'_{21}\big|\sqrt{p(1|1,1)}\sqrt{\sum_n\big|\gamma_{001n}+e^{i(\theta_3+\theta'_2)}\gamma_{111n}\big|^2}\\
&=(C_{30}C'_{20}\sqrt{\sum_n\big|\sqrt{p(1|0,0)}\gamma_{001n}+\sqrt{p(1|1,1)}e^{i(\theta_3+\theta'_2)}\gamma_{111n}\big|^2}-\big|C_{30}C'_{20}-C_{31}C'_{21}\big|\sqrt{p(1|1,1)})^2.\\
\end{aligned}
\end{equation}
Therefore, we obtain
\begin{equation}
\label{eqn-max-exp1-A}
\begin{aligned}
&\frac{\sum_n\big|\sqrt{p(1|0,0)}\gamma_{001n}+\sqrt{p(1|1,1)}e^{i(\theta_3+\theta'_2)}\gamma_{111n}\big|^2}{2(p(1|0,0)+p(1|1,1)+p(1|0,1)+p(1|1,0))}\leqslant\\
&\begin{cases}
&\frac{\big(\sqrt{p(1|3,2)}+\sqrt{p(1|0,1)}C_{30}C'_{21}+\sqrt{p(1|1,0)}C_{31}C'_{20}+\sqrt{p(1|1,1)}\big|C_{30}C'_{20}-C_{31}C'_{21}\big|\big)^2}{ 2(p(1|0,0)+p(1|1,1)+p(1|0,1)+p(1|1,0))C^2_{30}C'^2_{20}},\ \text{if}\ C_{30}C'_{20}\ne 0\\
&1-e_b,\ \text{if}\ C_{30}C'_{20}= 0.
\end{cases}
\end{aligned}
\end{equation}
Furthermore, by the same way, we can obtain that
\begin{equation}
\label{eqn-max-exp2-A}
\begin{aligned}
&\frac{\sum_n\big|\sqrt{p(1|0,0)}\gamma_{001n}+\sqrt{p(1|1,1)}e^{i(\theta_3+\theta'_2)}\gamma_{111n}\big|^2}{2(p(1|0,0)+p(1|1,1)+p(1|0,1)+p(1|1,0))}\leqslant\\
&\begin{cases}
&\frac{\big(\sqrt{p(1|3,2)}+\sqrt{p(1|0,1)}C_{30}C'_{21}+\sqrt{p(1|1,0)}C_{31}C'_{20}+\sqrt{p(1|0,0)}\big|C_{30}C'_{20}-C_{31}C'_{21}\big|\big)^2}{ 2(p(1|0,0)+p(1|1,1)+p(1|0,1)+p(1|1,0))C^2_{31}C'^2_{21}},\ \text{if}\ C_{31}C'_{21}\ne 0\\
&1-e_b,\ \text{if}\ C_{31}C'_{21}= 0.
\end{cases}
\end{aligned}
\end{equation}
Combining constraints \eqref{eqn-max-exp1-A} and \eqref{eqn-max-exp2-A}, we have
\begin{equation}
\label{eqn-max-exp-A}
\begin{aligned}
&\frac{\sum_n\big|\sqrt{p(1|0,0)}\gamma_{001n}+\sqrt{p(1|1,1)}e^{i(\theta_3+\theta'_2)}\gamma_{111n}\big|^2}{2(p(1|0,0)+p(1|1,1)+p(1|0,1)+p(1|1,0))}\\
&\leqslant max_{C,C'}f(C,C')\triangleq \varepsilon,
\end{aligned}
\end{equation}
in which $max_{C,C'}f(C,C')$ means searching over all $C_{30}$, $C_{31}$, $C'_{20}$ and $C'_{21}$ satisfying possible constraints to find the maximum value of function $f(C,C')$. Concretely, the $f(C,C')$ is given by
\begin{equation}
\label{eqn-f-A}
\begin{aligned}
&f(C,C')=\\
&\begin{cases}
&min\{\frac{\big(\sqrt{p(1|3,2)}+\sqrt{p(1|0,1)}C_{30}C'_{21}+\sqrt{p(1|1,0)}C_{31}C'_{20}+\sqrt{p(1|1,1)}\big|C_{30}C'_{20}-C_{31}C'_{21}\big|\big)^2}{ 2(p(1|0,0)+p(1|1,1)+p(1|0,1)+p(1|1,0))C^2_{30}C'^2_{20}},\\
&\frac{\big(\sqrt{p(1|3,2)}+\sqrt{p(1|0,1)}C_{30}C'_{21}+\sqrt{p(1|1,0)}C_{31}C'_{20}+\sqrt{p(1|0,0)}\big|C_{30}C'_{20}-C_{31}C'_{21}\big|\big)^2}{ 2(p(1|0,0)+p(1|1,1)+p(1|0,1)+p(1|1,0))C^2_{31}C'^2_{21}}\},\ \text{if}\ C_{30}C'_{20}\ne 0\ \text{and}\  C_{31}C'_{21}\ne 0\\
&\frac{\big(\sqrt{p(1|3,2)}+\sqrt{p(1|0,1)}C_{30}C'_{21}+\sqrt{p(1|1,0)}C_{31}C'_{20}+\sqrt{p(1|1,1)}\big|C_{30}C'_{20}-C_{31}C'_{21}\big|\big)^2}{ 2(p(1|0,0)+p(1|1,1)+p(1|0,1)+p(1|1,0))C^2_{30}C'^2_{20}},\ \text{if}\ C_{30}C'_{20}\ne 0\ \text{and}\ C_{31}C'_{21}= 0\\
&\frac{\big(\sqrt{p(1|3,2)}+\sqrt{p(1|0,1)}C_{30}C'_{21}+\sqrt{p(1|1,0)}C_{31}C'_{20}+\sqrt{p(1|0,0)}\big|C_{30}C'_{20}-C_{31}C'_{21}\big|\big)^2}{ 2(p(1|0,0)+p(1|1,1)+p(1|0,1)+p(1|1,0))C^2_{31}C'^2_{21}},\ \text{if}\ C_{30}C'_{20}= 0\ \text{and}\ C_{31}C'_{21}\ne 0\\
&1-e_b,\ \text{if}\  C_{30}C'_{20}= 0\ \text{and}\ C_{31}C'_{21}= 0, \\
\end{cases}
\end{aligned}
\end{equation}
where $min\{a,b\}$ yields the smaller one of real numbers $a$ and $b$

Next we try to calculate above bounds by considering some constraints on it. Since we have known $p(1|3,0)$, $p(1|3,1)$ and other probabilities for mismatched basis. In other words, we have learned $\big|_M\langle1|U_{Eve}|\varphi_3\rangle_C|\varphi'_0\rangle_D\big|^2$, $\big|_M\langle1|U_{Eve}|\varphi_3\rangle_C|\varphi'_1\rangle_D\big|^2$, $\big|_M\langle1|U_{Eve}|\varphi_0\rangle_C|\varphi'_2\rangle_D\big|^2$, $\big|_M\langle1|U_{Eve}|\varphi_1\rangle_C|\varphi'_2\rangle_D\big|^2$. Substitute the relations \eqref{eqn-varphi23} and Eq. \eqref{eqn-model1} into these equations, we have

\begin{equation}
\label{eqn-constraints3-A}
\begin{aligned}
&C_{30}^2p(1|0,0)+C_{31}^2p(1|1,0) + 2 C_{30} C_{31}\sqrt{p(1|0,0)p(1|1,0)}Re\{e^{i\theta_3} \langle\Gamma_{001}|\Gamma_{101}\rangle\}=p(1|3,0)\\
&C_{30}^2p(1|0,1)+C_{31}^2p(1|1,1) + 2 C_{30} C_{31}\sqrt{p(1|0,1)p(1|1,1)}Re\{e^{i\theta_3} \langle\Gamma_{011}|\Gamma_{111}\rangle\}=p(1|3,1)\\
&C'^2_{20}p(1|0,0)+C'^2_{21}p(1|0,1) + 2 C'_{20} C'_{21}\sqrt{p(1|0,0)p(1|0,1)}Re\{e^{i\theta'_2} \langle\Gamma_{001}|\Gamma_{011}\rangle\}=p(1|0,2)\\
&C'^2_{20}p(1|1,0)+C'^2_{21}p(1|1,1) + 2 C'_{20} C'_{21}\sqrt{p(1|1,0)p(1|1,1)}Re\{e^{i\theta'_2} \langle\Gamma_{101}|\Gamma_{111}\rangle\}=p(1|1,2),\\
\end{aligned}
\end{equation}
where $Re\{x\}$ returns the real part of a complex number $x$. For the ease of numerical computation, we rewrite above constraints as

\begin{equation}
\label{eqn-constraints-A}
\begin{aligned}
&-2\sqrt{p(1|0,0)p(1|1,0)}C_{30}C_{31}\leqslant p(1|3,0)-p(1|0,0)C^2_{30}-p(1|1,0)C^2_{31}\leqslant 2\sqrt{p(1|0,0)p(1|1,0)}C_{30}C_{31}\\
&-2\sqrt{p(1|0,1)p(1|1,1)}C_{30}C_{31}\leqslant p(1|3,1)-p(1|0,1)C^2_{30}-p(1|1,1)C^2_{31}\leqslant 2\sqrt{p(1|0,1)p(1|1,1)}C_{30}C_{31}\\
&-2\sqrt{p(1|0,0)p(1|0,1)}C'_{20}C'_{21}\leqslant p(1|0,2)-p(1|0,0)C'^2_{20}-p(1|0,1)C'^2_{21}\leqslant 2\sqrt{p(1|0,0)p(1|0,1)}C'_{20}C'_{21}\\
&-2\sqrt{p(1|1,0)p(1|1,1)}C'_{20}C'_{21}\leqslant p(1|1,2)-p(1|1,0)C'^2_{20}-p(1|1,1)C'^2_{21}\leqslant 2\sqrt{p(1|1,0)p(1|1,1)}C'_{20}C'_{21}.\\
\end{aligned}
\end{equation}

Now the task is to calculate $\varepsilon$ in Eq.\eqref{eqn-max-exp-A} with constraints Eqs. \eqref{eqn-constraints-A}. Then $e_p$ can be calculated easily. Therefore, the secure-key rate is given by
\begin{equation}
\label{eqn-key-rate-formula1}
R=1-H(e_b)-H(e_p)=1-H(e_b)-H(min\{\varepsilon+e_b,1/2\}),
\end{equation}
where $H(x)=-x \log x -(1-x) \log (1-x)$ is the Shannon's binary entropy function.
Note that by employing the recently developed security proofs \cite{Caves:deFinetti:2002,Renner:deFinetti:2009}, our protocol can be secure against the most general attacks, although the attack analyzed here is collective attack.

%\bibitem{Biham}
%Eli Biham, Bruno Huttner, Tal Mor, Phys. Rev. A \textbf{54}, 2651 (1996).
%\bibitem{Inamori}
%Hitoshi Inamori, Algorithmica 34, 340 (2002).
%\bibitem{MDIQKD2}
%Kiyoshi Tamaki, Hoi-Kwong Lo, Chi-Hang Fred Fung, and Bing Qi , Phys. Rev. A \textbf{85}, 042307 (2012).
%\bibitem{wang}
%Xiang-Bin Wang, arXiv:1207.0392.
%\bibitem{semi}
%Marcin Pawlowski and Nicolas Brunner, Phys. Rev. A \textbf{84}, 010302 (2011).
%
%\bibitem{qdf} C. M. Caves, C. A. Fuchs and R. Schack, J.
%Math. Phys. \textbf{43}, 4537 (2002); R. K\"{o}nig and R. Renner, J.
%Math. Phys. \textbf{46}, 122108 (2005).
%
%\bibitem{renner} R. Renner, Nature Phys. \textbf{3}, 645 (2007).
%
%\bibitem{ckr} M. Christandl, R. K\"{o}nig, and R. Renner,
%Phys. Rev. Lett. \textbf{102}, 020504 (2009).

%%%%%%%%%%%%%%%%%%%%%%%%%%%%%%%%%%%%%%%%
% choose a style
%\bibliographystyle{ieeetr}
%\bibliographystyle{unsrt}
\bibliographystyle{apsrev4-1}
%%%%%%%%%%%%%%%%%%%%%%%%%%%%%%%%%%%%%%%%

%%%%%%%%%%%%%%%%%%%%%%%%%%%%%%%%%%%%%%%%
% choose a .bib file
\bibliography{Biblisource}
%%%%%%%%%%%%%%%%%%%%%%%%%%%%%%%%%%%%%%%%

\end{document}